# Extended Theory of Harmonic Maps Connects General Relativity to Chaos and Quantum Mechanism

Gang Ren[1,*] and Yi-Shi Duan[2,*,#]

[1] The Molecular Foundry, Lawrence Berkeley National Laboratory, Berkeley CA 94720, USA
[2] The Department of Physics, Lanzhou University, Lanzhou, Gansu 730000, China

* To whom correspondence should be addressed: G. Ren, Lawrence Berkeley National Laboratory, the Molecular Foundry, 1 Cyclotron Road, Berkeley, CA 94720, USA, E-mail: gren@lbl.gov or Y.S. Duan, The Department of Physics, Lanzhou University, Lanzhou, Gansu 730000, China, E-mail: ysduan@lzu.edu.cn

[#] Deceased, December, 21, 2016

**Key Words**: Extended harmonic maps, extended theory of harmonic maps, chaos equation, general relativity, quantum physics

**Abstract**
General relativity and quantum mechanism are two separate rules of modern physics explaining how nature works. Both theories are accurate, but the direct connection between two theories was not yet clarified. Recently, researchers blur the line between classical and quantum physics by connecting chaos and entanglement. Here, we showed the early reported extended HM theory that included the general relativity can also be used to recover the classic chaos equations and even the Schrödinger equation in quantum physics, suggesting the extended theory of harmonic maps may act as a universal theory of physics.

## Introduction

Harmonic map (HM) has been used to recover the Ernst formulation of Einstein's equations in general relativity [1-4], in which the solution of Euler-Lagrange equations of the harmonic maps could be obtained through solving the Laplace-Beltrami's equation and the geodesic equations [3]. From the viewpoint of physics, the traditional theory of harmonic maps contains only the kinetic energy term. For a wider application of this method, the Lagrangian of the traditional HM theory was supplemented with a potential energy term by Duan *et. al.* [5]. This extended HM theory is helpful for studying the travelling wave or soliton solutions of some types of nonlinear partial differential equations. Recently, researchers discovered the classical chaos showed a connection to quantum physics via entanglement [6,7]. Here, we asked whether the extended HM theory can recover the classical chaos questions even the Schrödinger equation in quantum physics.

In classic physics, all physical processes are described by differential equations. Over the last several decades, many nonlinear ordinary differential equations possessing chaotic behaviours have been discovered[8], including the nonlinear equation for anharmonic system in periodic fields [9] given by

$$\frac{d^2x}{dt^2} + k\frac{dx}{dt} - \beta x + \alpha x^3 = b Cos(\varpi t) \ , \quad (1)$$

where k, α and β are control parameters, and equation for a parametrically excited pendulum given by[10]

$$\frac{d^2x}{dt^2} + k\frac{dx}{dt} + [\alpha + \beta Cos(\varpi t)] Sin(x) = 0, \quad (2)$$

where k, α and β are control parameters. To date, the study of chaotic behaviours using the space distribution of nonlinear partial differential equations has been rare and is currently only in its initial stage.

Here, we used the extended theory of HM derivate the classical chaos equations (1) and (2) and even the Schrödinger equation for a one-dimensional harmonic oscillator. Considering the extended theory of HM can cover both of the Einstein's equations and Schrödinger equation may function as a fundamental rule our universe.

## The extended theory of harmonic maps (HM)

The theory of HM [11-13] became an important branch of mathematical physics decades ago and has been applied to a wide variety of problems in mathematics and theoretical physics. In this section, we re-introduce the formulation of the extended theory of HM that has been reported by Duan *et. al.* [5].

Let M and N be two Riemannian manifolds with local coordinates $x^\mu$ ($\mu = 1,2,...,m$) on M and local coordinates $\Phi^A$ (A = 1,2,...,n) on N. The metrics on M and N are denoted by

$$dl^2 = g_{\mu\nu}(x)dx^\mu dy^\nu \quad ; \quad \dim(M)=m$$
$$dL^2 = G_{AB}(\Phi)d\Phi^A d\Phi^B \ ; \quad \dim(N)=n \quad (3)$$

respectively. A mapping

$$\Phi: \ M \ \rightarrow \ N$$
$$x \ \rightarrow \ \Phi(x)$$

is called an extended harmonic map if it satisfies the Euler-Lagrange equation resulting from the variational principle $\delta I = 0$, using the action

$$I = \int d^n x \sqrt{g}[-\frac{1}{2}g^{\mu\nu}\partial_\mu \Phi^A \partial_\nu \Phi^B G_{AB}(\Phi) + V(\Phi)] \tag{4}$$

where $\partial_\mu = \frac{\partial}{\partial x^\mu}$, $V(\Phi)$ is the potential function of $\Phi^A$, and $V(\Phi) = V(\Phi^1, \Phi^1, ...., \Phi^n)$.

It is obvious that the traditional HM corresponds to the case of $V(\Phi) = 0$ in equation (4). It has been showed that, in two dimensional case, the solutions of the Ernst equation in Einstein's general relativity can be recovered from the Euler's equations of traditional HM [3,14].

The conditions for a map to be harmonic are given by the Euler-Lagrange equations

$$\frac{\partial L}{\partial \Phi^A} - \partial_\mu \frac{\partial L}{\partial \partial_\mu x \Phi^A} = 0; \quad A = 1,2,...,n \tag{5}$$

where

$$L = -\frac{1}{2}g^{\mu\nu}\partial_\mu \Phi^A \partial_\nu \Phi^B G_{AB}(\Phi)\sqrt{g} + V(\Phi)\sqrt{g} \tag{6}$$

The equation (5) follows the notational conventions of Wald's equation[15]. This equation (5) appeared in the original paper of Duan (as equations 2)[5].

By substituting (6) into (5), we can obtain the Euler-Lagrange equations of extended HM given by

$$\frac{1}{\sqrt{g}}\partial_\mu(\sqrt{g}g^{\mu\nu}\partial_\nu \Phi^A) + \Gamma^A_{BC}\partial_\mu \Phi^B \partial_\nu \Phi^C g^{\mu\nu} + G^{AB}\frac{\partial V(\Phi)}{\partial \Phi^B} = 0, \tag{7}$$

where $\Gamma^A_{BC}$ are the Christoffels symbols on manifold N given by

$$\Gamma^A_{BC} = \frac{1}{2}G^{AD}[\frac{\partial G_{BD}}{\partial \Phi^C} + \frac{\partial G_{CD}}{\partial \Phi^B} - \frac{\partial G_{BC}}{\partial \Phi^D}]. \tag{8}$$

In order to obtain a special type of solution of partial differential equation (7), Duan used his published approach [3], which could be especial convenient to be used to study the soliton solution of the partial differential equation (7). In brief, in the case of $\Phi^A$ (A = 1,2,...,n) are functions solely of the argument σ, and σ is a function of $x^\mu$ on the manifold M:

$$\Phi^A = \Phi^A(\sigma); \tag{9}$$
$$\sigma = \sigma(x). \tag{10}$$

The equation (7) can be written as following,

$$\frac{1}{\sqrt{g}}\partial_\mu(\sqrt{g}g^{\mu\nu}\partial_\nu \sigma)\frac{d\Phi^A}{d\sigma} + G^{AB}\frac{\partial V(\Phi)}{\partial \Phi^B} + (\frac{d^2\Phi^A}{d\sigma^2} + \Gamma^A_{BC}\frac{d\Phi^B}{d\sigma}\frac{d\Phi^C}{d\sigma})g^{\mu\nu}\partial_\mu \sigma \partial_\nu \sigma = 0. \tag{11}$$

This important equations (11) appear in the original paper of Duan (as equation 7)[5].

The soliton solution, such as the solution of Sine-Gorden equation, and the travelling wave solution can be found in certain cases of the metrics and potential function in the extended Euler-Lagrange equation (11).

If the function $\sigma = \sigma(x)$ satisfies the Laplace-Beltrami equations

$$\frac{1}{\sqrt{g}}\partial_\mu(\sqrt{g}g^{\mu\nu}\partial_\nu\sigma) = 0, \tag{12}$$

the equation (11) will take the form

$$\frac{d^2\Phi^A}{d\sigma^2} + \Gamma^A_{BC}\frac{d\Phi^B}{d\sigma}\frac{d\Phi^C}{d\sigma} = -\frac{1}{f}G^{AB}\frac{\partial V(\Phi)}{\partial \Phi^B}, \tag{13}$$

where

$$f = g^{\mu\nu}\partial_\mu\sigma\partial_\nu\sigma \tag{14}$$

is a scalar function on the manifold M.

Because $f$ is not the function of $\Phi^A$, a rescaling of $V(\Phi)$ as $V(\Phi) = fU(\Phi)$ re-expresses the equation (13) as

$$\frac{d^2\Phi^A}{d\sigma^2} + \Gamma^A_{BC}\frac{d\Phi^B}{d\sigma}\frac{d\Phi^C}{d\sigma} = -G^{AB}\frac{\partial U(\Phi)}{\partial \Phi^B}. \tag{15}$$

Equation (15) can be represented as the geodesic equation of a particle on the Riemannian manifold N subjected to an external force of

$$F^A = -G^{AB}\frac{\partial U(\Phi)}{\partial \Phi^B}. \tag{16}$$

This equation (15) will become equivalent to the usual geodesic equations on the manifold N if the external force $F^A = 0$ $(A = 1,2,\ldots,n)$. Equation (15) is physical dynamical equation.

**Chaotic solutions of the extended HM equations**

If the chaos equation could be derivate from the equation (15) under certain cases of metrics and potential function, the chaotical behaviour satisfied the extended HM equations. In order to evidence whether the chaotic behaviours exist in the partial differential equation (15), we simplified the case as that the M is the pseudo-Euclidean space-time and N is a 2-dimensional manifold with coordinates $\{\Phi^1, \Phi^2\}$. We suppose that $\Phi^1$ and $\Phi^2$ are functions of an argument $\sigma = \sigma(x) = k_\mu x^\mu$ and further assume that $\Phi^2 = \sigma$, i.e.,

$$\Phi^1 = \Phi(\sigma); \quad \Phi^2 = \sigma; \quad \sigma = k_\mu x^\mu. \tag{17}$$

Under this simplified case, the equation (15) can be expressed as

$$\frac{d^2\Phi}{d\sigma^2} + \Gamma^1_{11}\left(\frac{d\Phi}{d\sigma}\right)^2 + 2\Gamma^1_{12}\frac{d\Phi}{d\sigma} + \Gamma^1_{12} = -\left(G^{11}\frac{\partial U(\Phi,\sigma)}{\partial \Phi} + G^{12}\frac{\partial U(\Phi,\sigma)}{\partial \sigma}\right); \tag{18}$$

$$\Gamma^2_{11}\left(\frac{d\Phi}{d\sigma}\right)^2 + 2\Gamma^2_{12}\frac{d\Phi}{d\sigma} + \Gamma^2_{22} = -\left(G^{21}\frac{\partial U(\Phi,\sigma)}{\partial \Phi} + G^{22}\frac{\partial U(\Phi,\sigma)}{\partial \sigma}\right). \tag{19}$$

Eliminating the term $(\frac{d\Phi}{d\sigma})^2$ from the differential equations (18) and (19), we obtain

$$\frac{d^2\Phi}{d\sigma^2} + 2\left(\Gamma^1_{12} - \frac{\Gamma^1_{11}\Gamma^2_{12}}{\Gamma^2_{11}}\right)\frac{d\Phi}{d\sigma} + \left(\Gamma^1_{22} - \frac{\Gamma^1_{11}\Gamma^2_{22}}{\Gamma^2_{11}}\right)$$
$$+ \left[G^{11}\frac{\partial U(\Phi,\sigma)}{\partial \Phi} + G^{12}\frac{\partial U(\Phi,\sigma)}{\partial \sigma} - \frac{\Gamma^1_{11}}{\Gamma^2_{11}}\left(G^{21}\frac{\partial U(\Phi,\sigma)}{\partial \Phi} + G^{22}\frac{\partial U(\Phi,\sigma)}{\partial \sigma}\right)\right] = 0$$
(20)

If the chaos equation can be derivate from these geodesical ordinary differential equations (20), the partial differential equation (15) must contain the chaotic behaviours.

By given the metrics on manifold N are diagonal and take the following form:

$$G_{11} = e^{\Phi+\sigma}; \quad G_{12} = 0; \quad G_{21} = 0; \quad G_{22} = (k-1)e^{\Phi+\sigma}, \tag{21}$$

where k is a constant.

The Christoffels symbols (8) on the 2-dimensional manifold are calculated as following,

$$\Gamma^1_{11} = \tfrac{1}{2}; \quad \Gamma^1_{12} = \tfrac{1}{2}; \quad \Gamma^1_{22} = -\tfrac{1}{2}(k-1); \quad \Gamma^2_{11} = -\tfrac{1}{2(k-1)}; \quad \Gamma^2_{12} = \tfrac{1}{2}; \quad \Gamma^2_{22} = \tfrac{1}{2}. \tag{22}$$

The detailed procedures of calculation showed in the supplementary information.

By substituting (21) and (22) into (20), the differential equation (20) can be written as following,

$$\frac{d^2\Phi}{d\sigma^2} + k\frac{d\Phi}{d\sigma} + (e^{-\Phi-\sigma})\left[\frac{\partial U(\Phi,\sigma)}{\partial \Phi} + \frac{\partial U(\Phi,\sigma)}{\partial \sigma}\right] = 0. \tag{23}$$

We found, under following given potential function U(Φ, σ), i.e.

$$U(\Phi,\sigma) = e^{\Phi+\sigma}\left[\tfrac{1}{2}\alpha\Phi^3 - \tfrac{3}{4}\alpha\Phi^2 - \tfrac{1}{4}(2\beta-3\alpha)\Phi + \tfrac{1}{8}(2\beta-3\alpha) - \frac{2b}{\omega^2+4}\cos(\omega\sigma) - \frac{\omega b}{\omega^2+4}\sin(\omega\sigma)\right], \tag{24}$$

Substituting (24) into equation (23), we obtain

$$\frac{d^2\Phi}{d\sigma^2} + k\frac{d\Phi}{d\sigma} - \beta\Phi + \alpha\Phi^3 = b\cos(\omega\sigma); \quad \sigma = k_\mu x^\mu \tag{25}$$

The detailed procedures of calculation showed in the supplementary information.

Comparing the equation (25) with equation (1), i.e., the equation of anharmonic system in a periodic field, we find that they are of the same form. Since the equation (1) possesses chaotic states that are characterized by the existence of a strange attractor in the phase space, this implies that equation (23) for $\Phi = \Phi(\sigma)$ should possess the same chaotic profile.

By giving a new potential function $U = U(\Phi, \sigma)$, i.e.

$$U(\Phi, \sigma) = e^{\Phi+\sigma} \{\frac{2}{5}\alpha \sin(\Phi) - \frac{1}{5}\alpha\cos(\Phi) + \frac{\beta}{25+6\omega^2+\omega^4}[\omega(\omega^2 + 3)\sin(\omega\sigma)\sin(\Phi) - 4\omega\sin(\omega\sigma)\cos(\Phi) + 2(\omega^2 + 5)\cos(\omega\sigma)\sin(\Phi) + (\omega^2 - 5)\cos(\omega\sigma)\cos(\Phi)]\},$$
(26)

in which, $\alpha$ and $\beta$ are constants, the equation (23) become following,

$$\frac{d^2\Phi}{d\sigma^2} + k\frac{d\Phi}{d\sigma} + [\alpha + \beta \cos(\omega\sigma)]\sin(\Phi) = 0; \quad \sigma = k_\mu x^\mu \quad (27)$$

The detailed procedures of calculation showed in the supplementary information.

Comparing the equation (27) with equation (2), *i.e.*, the nonlinear equation of a parametrically excited damped pendulum, we find that they are of the same form. Since the equation (2) possesses chaotic states that are characterized by the existence of a strange attractor in the phase space, this implies that equation (27) for $\Phi(\sigma)$ should possess the same chaotic profile.

**Quantum mechanism solutions of the extended HM equations**

The extended equation of HM contains the chaotic solution, which recently found with connection to quantum physics via entanglement [1,2]. It would be interested to test whether the Schrödinger equation in quantum physics can be derived. As a simple case, we studied the Schrödinger equation for a one-dimensional harmonic oscillator as below,

$$\frac{d^2\Psi}{dx^2} = -\frac{2m}{\hbar^2}\left(E - \frac{1}{2}Kx^2\right)\Psi(x), \quad (28)$$

where $\Psi(x)$ is the wave function, $E$ is the total energy (constant), and $K$ is the force constant (the force on the mass being $F = -Kx$, proportional to the displacement $x$ and directed towards the origin.

By given the metrics on manifold N are diagonal and take the following form:

$$G_{11} = e^{\Phi+\sigma}; \quad G_{12} = 0; \quad G_{21} = 0; \quad G_{22} = -e^{\Phi+\sigma}, \quad (29)$$

This equation (29) is a special case of equation (21), *i.e.* $k = 0$. Thus, the differential equation (20) can be written as following,

$$\frac{d^2\Phi}{d\sigma^2} + (e^{-\Phi-\sigma})\left[\frac{\partial U(\Phi,\sigma)}{\partial \Phi} + \frac{\partial U(\Phi,\sigma)}{\partial \sigma}\right] = 0. \quad (30)$$

By given the potential function $U(\Phi, \sigma)$ as following, *i.e.*

$$U(\Phi, \sigma) = e^{\Phi+\sigma}\left[-\frac{m}{2\hbar^2}K\Phi\sigma^2 - \frac{m}{4\hbar^2}K\sigma^2 - \frac{m}{2\hbar^2}K\Phi\sigma + \frac{2m}{\hbar^2}(E-K)\Phi - \frac{m}{2\hbar^2}K\sigma + \frac{m}{\hbar^2}\left(\frac{3}{4}K - E\right)\right], \quad (31)$$

where $E, K, m, \hbar$ are the constants,

Substituting (31) into equation (30), we obtain

$$\frac{d^2\Phi}{d\sigma^2} = -\frac{2m}{\hbar^2}\left(E - \frac{1}{2}K\sigma^2\right)\Phi(\sigma), \tag{32}$$

The detailed procedures of calculation showed in the supplementary information.

Comparing the equation (32) with equation (28), we find that they are of the same form when $\Phi(\sigma) = \Psi(x)\ and\ \sigma = x$. Since the equation (28) is one of typical equation of Schrodinger equation in quantum physics, this implies that the equation (15) contains the Schrodinger equation in quantum physics.

Since the extended HM (when the potential function $V(\Phi) = 0$ in the equation 4) can be recovered the Einstein's equations in general relativity [14], and the same theory can also recover the Schrodinger equation in quantum physics, we propose here the extended equation of HM as a single and universal theory for our universe.

**Acknowledgments:** We thank Mr. Wencheng Ren to preserve the original results and manuscript. Work at the Molecular Foundry was supported by the Office of Science, the Office of Basic Energy Sciences, and the U.S. Department of Energy, under Contract No. DE-AC02-05CH11231.

**Author Contributions:** This project was initiated by GR and YD., GR calculated the solution, and YS validated the solution. GR drafted the initial manuscript, which was revised by YD.

**Additional Information**

**Competing financial interests**
The author(s) declare no competing financial interests.